# OpenHEC: A Framework for Application Programmers to Design FPGA-based Systems

Zhilei Chai, Zhibin Wang, Wenmin Yang, Shuai Ding and Yuanpu Zhang
State Key Laboratory of Mathematical Engineering and Advanced Computing, Wuxi, China, 214125
School of IoT Engineering, Jiangnan University, Wuxi, China, 214122
Email: {zlchai,stepzhibin, ouyang.wm, sding, yuanpzh}@jiangnan.edu.cn

*Abstract*—Today, there is a trend to incorporate more intelligence (e.g., vision capabilities) into a wide range of devices, which makes high performance a necessity for computing systems. Furthermore, for embedded systems, low power consumption should be generally considered together with high computing performance. FPGAs, as programmable logic devices able to support different types of fine-grained parallelisms, their power and performance advantages were recognized widely. However, designing applications on FPGA-based systems is traditionally far from a task can be carried out by software programmers. Generally, hardware engineers and even system-level software engineers have more hardware/architectural knowledge but fewer algorithm and application knowledge. Thus, it is critical for computing systems to allow application-level programmers to realize their idea conveniently, which is popular in computing systems based on the general processor. In this paper, the OpenHEC (Open Framework for High-Efficiency Computing) framework is proposed to provide a design framework for application-level software programmers to use FPGA-based platforms. It frees users from hardware and architectural details to let them focus more on algorithms/applications. This framework was integrated with the commercial Xilinx ISE/Vivado to make it to be used immediately. After implementing a widely-used feature detection algorithm on OpenHEC from the perspective of software programmers, it shows this framework is applicable for application programmers with little hardware knowledge.

## I. INTRODUCTION

As we know, in the recent decade, manufactures have given up attempting to extract more performance from a single core with higher frequency, most opportunities for performance growth in mainstream general-purpose computers have been tied to their exploitation of the increasing number of processor cores. Many topics related to multi-core research have been investigated [1] [2]. However, as introduced in [2], current parallel architectures allow good speedups on regular programs such as dense-matrix type programs, but are mostly handicapped on irregular programs such as computer vision programs.

FPGAs, as programmable logic devices able to support different types of fine-grained parallelisms, are good at processing irregular programs. Their power and performance benefits were frequently proven [3] [4]. However, nowadays implementing applications on pure FPGAs is mostly an engineering discipline carried out by highly trained specialists due to low-level languages such as VHDL/Verilog HDL are generally used and hardware details have to be handled. Furthermore, non-trivial driver issues have to be usually dealt with before user applications can be implemented.

Recently, there appear heterogeneous SoCs for FPGA such as Xilinx Zynq-7000 [5] and Altera SoC FPGA [6]. One advantage of the heterogeneous FPGA is that its general processor part is able to free FPGA users from dealing with driver issues to a large extent. The other advantage is that the heterogeneous FPGA has potential to achieve better energy efficiency by combining traditional processors with unconventional cores, which has been studied theoretically [8] and practically [9]. On the other hand, HLS (High-Level-Synthesis) [7] technology today is mature enough to support C/C++ with higher abstraction for FPGA designing. For example, Xilinx released the Vivado HLS [10] to support programmers to use high-level language such as C/C++ to design systems based on Xilinx FPGAs. The Zynq SoC coupled with HLS construct a noticeable development choice today. However, although the heterogeneous SoC coupled with HLS improves description abstraction of system designing and frees users from non-trivial FPGA driver issues, it is more suitable for system-level programmers. It is because the programmer has to grasp knowledge of algorithm description, architecture and interfaces, physical address allocation, software drivers implementating and application software implementing. The drawback of this design flow is that system-level programmers generally have more architectural knowledge about hardware and software but fewer knowledge about algorithms and applications, and vise versa.

HLS for FPGA designing can be viewed as the compiler in general-purpose computer systems. Obviously, HLS users still need to face the bare FPGA although high-level languages are used. Thus, similar to the operating system in general-purpose computers, a virtual layer is also necessary for FPGAs to improve development efficiency through shielding hardware details further from users. Some work have been conducted already from this perspective. The LEAP [12] is an FPGA operating system addressing latency-insensitive design, communication abstractions and memory models for multi-FPGA and hybrid algorithms. The RAMP [1] project aims to provide a FPGA-based research accelerator for multiple processors though FPGA virtualization. The CoRAM [13] project focuses on application interfaces to memory hierarchies and memory models for reconfigurable logic. More related work can be found in [11] [14] [15] etc.

In this paper, a framework called OpenHEC is proposed for application programmers to use FPGA-based heterogeneous platforms. This framework is optimized for computer vision applications in embedded systems. Rather than implementing a set of development tools from scratch, OpenHEC is integrated into the commercial Xilinx Vivado to make it to be used immediately. For programmers using OpenHEC to design applications, C/C++ are used for both the CPU part and the FPGA part. The main entry of the program is on the CPU part and the FPGA part is reconfigured and invoked as necessary during execution. OpenHEC frees users from handling hardware and architecture details to allow them to focus more on algorithms/applications. The framework was verified by implementing a popular computer vision algorithm for feature detection, the SURF (Speeded-Up Robust Features) detection[16], from the perspective of an application-level software programmer. It shows this framework is applicable for application programmers to use FPGA-based heterogeneous platforms.

## II. OPENHEC ARCHITECTURE

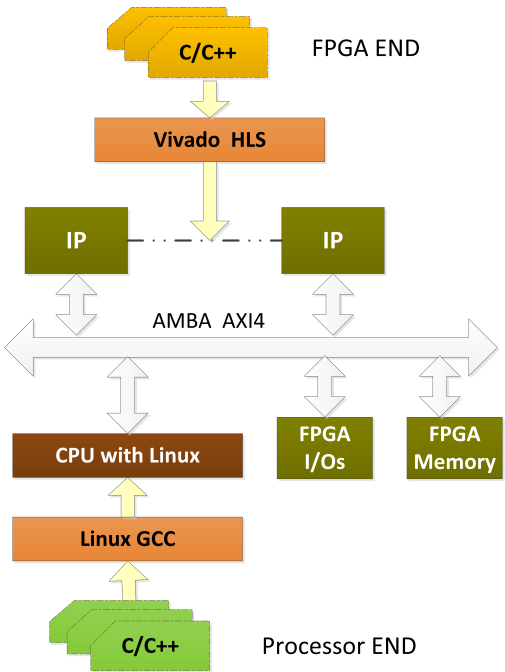

Fig. 1. Architecture of the OpenHEC Framework

As shown in Fig. 1, the OpenHEC framework provides support for application programmers using C/C++ on both the FPGA end and the processor end. At the FPGA end, algorithms or modules to be accelerated by hardware parallelism are described in C/C++, which are designed with an interface compliant with AMBA (Advanced Microcontroller Bus Architecture) AXI (Advanced eXtensible Interface). Then, they are translated and synthesized by HLS tools to generate IP (Intellectual Property) cores. These AXI-compliant IPs can be mounted to suited locations to construct the whole FPGA part. User's IPs developed by application programmers are application-oriented. Other modules such as FPGA driven I/Os and memories are provided as part of the OpenHEC framework. Thus, application programmers are able to pay all attention to how to implement their application-specific algorithms/modules best. Due to the C/C++ language supported and optimizing schemes provided in Vivado HLS, algorithms/modules are possible to be developed by software programmers. Through allocating fixed addresses in advance to avoid users to mount IPs manually in different implementations, it further makes C/C++ application programmers without much architectural knowledge able to develop algorithms for FPGAs. At the processor end, algorithms/modules to be executed on CPUs are also developed with C/C++. They are compiled and linked to generate executable files for CPUs equipped with operating systems. Modules executed on CPUs are able to communicate with that of FPGA part through the interface compliant with AXI.

### A. System Design of the FPGA Part

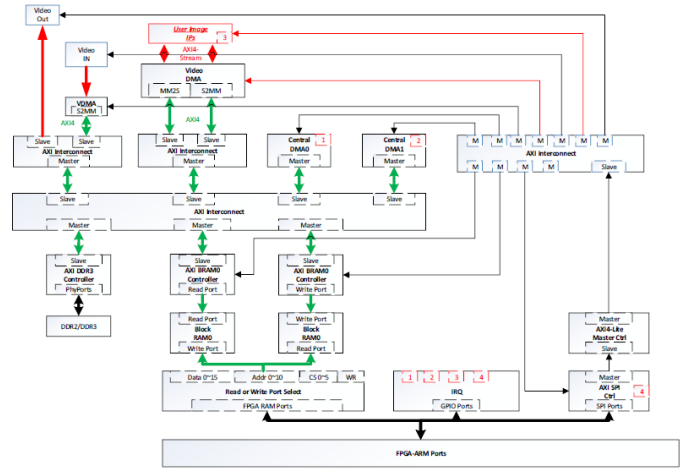

Fig. 2. Structure of the FPGA Part

As shown in Fig. 2, in the current version of the OpenHEC framework, AXI is used as the interconnection to connect IPs, I/Os, the FPGA memory and the processor. AXI4, AXI stream and AXI-lite are implemented within the FPGA. Through implementing this architecture and allocating addresses for user IPs in advance, it makes IPs, FPGA I/Os and memories in the FPGA part able to be 'seen' by the processor and be managed as resources. Allocated address space for IPs can be mapped to virtual addresses managed by operating system on the processor, it frees CPU users from developing device drivers to communicate with IPs. Thus, application programmers on CPU are qualified to develop algorithms on the processor end. For scalability, the OpenHEC framework is designed for containing more IPs and I/Os as necessary.

### B. System Design of the CPU Part

In order to facilitate application programmers on CPU to use functions implemented on the FPGA part, OpenHEC provides a series of APIs for them to use FPGA modules without implementing specific drivers, including: to configure FPGA bitstream with specific functional modules, to set parameters for FPGA modules, to communicate with the FPGA and so on.

$Config(lib\_name, bin\_file, algorithm\_info)$; To configure the FPGA with designated bitstream and save corresponding information about FPGA resources in algorithm_info for later management by the processor.

$Algorithm\_set(point * IPcore, parameter\_name, parameter\_value)$; To configure algorithm parameters for an FPGA module before working.

$FPGA\_mem\_request(point * point\_name, data\_size)$; $FPGA\_mem\_release(point * point\_name)$; To request and release a block of memory connected with the FPGA.

$Arm\_tx(point * ARM\_source, point\ FPGA\_destination, int\ ncols, int\ nrows)$; $ARM\_rx(point * ARM\_destination, point\ FPGA\_source, int\ ncols, int\ nrows)$; To do memory copying between memory blocks of the processor and the FPGA.

$Start(point * IPcore)$; To start FPGA modules after parameters configuring.

$reset(point * IPcore)$; To reset FPGA modules when necessary.

## III. DESIGN FLOW BASED ON OPENHEC FRAMEWORK

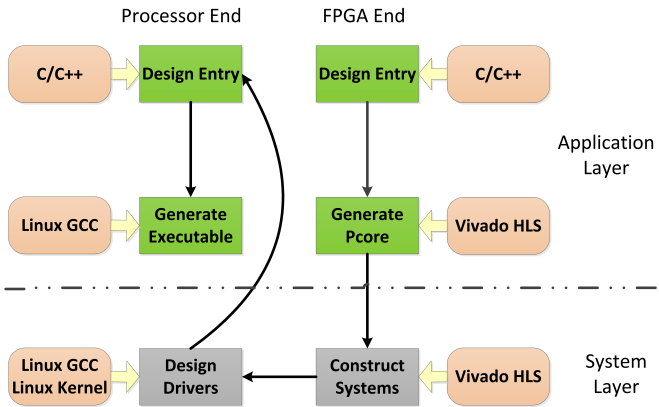

Fig. 3. Design Flow of FPGA-based Heterogeneous Systems with and without the OpenHEC Framework

Fig. 3 shows the design flow of FPGA-based heterogeneous systems for C/C++ programmers. As shown in Fig. 3, after the OpenHEC framework being employed, only the part above the dotted line is necessary for the design flow, which is introduced as follows.

**1) Design algorithms/modules for the FPGA part using C/C++.** In order to facilitate users to design AXI-compliant algorithms/modules, a template is provided in OpenHEC. Users can embed their design into this template and use the #pragma to construct AXI-compliant interface for it. For algorithm optimization to exploit more parallelism, Vivado HLS provides optimizing schemes. Users also can do optimization manually if they have some architectural knowledge. Please note that the simulation stage used in debugging provided in Vivado HLS is not discussed here.

**2) Generate IPs from algorithms designed above.** Algorithms and modules implemented by C/C++ can be processed by Vivado to generate an IP which can be used as a component later. Then, this IP can be connected with the AXI interconnection at a suitable address. Because a list of fixed addresses are allocated and the bus interconnection architecture is implemented in advance, users can connect their IPs to specific addresses without requiring for much architectural knowledge.

**3) Design algorithms/modules for the CPU part using C/C++.** Algorithms/modules suited to be executed on CPUs and the main procedure can be implemented using C/C++ in a conventional software development environment. As introduced in Sec. II-B, APIs are provided by OpenHEC to let users configure FPGA bitstream, set parameters and use FPGA modules. Thus, programmers are able to manipulate FPGA modules and parse data through using those APIs. Users designing a new FPGA module are responsible to construct another API for upper users to call FPGA function in pure software fashion.

**4) Generate executable file for CPUs.** Program above can be compiled and linked by Linux GCC to generate the executable file for CPUs. This executable file is able to reconfigure FPGA and invoke FPGA modules as necessary during execution.

Users of OpenHEC can be viewed as three different kinds as follows. a) Users to implement algorithms for both FPGAs and CPUs. b) Users to implement algorithms for FPGAs only. c) Users to implement algorithms for CPUs only. For the first kind of users, the whole design flow described above is employed. For the second kind of users, besides implementing IPs, they need provide a software API for upper users on CPUs which has specific function and can configure and manipulate corresponding IPs as necessary. For the third kind of users, they just use APIs provided by the second kind of users from the perspective of traditional software programmers on CPUs. They even do not know FPGA modules are employed. With more IPs being implemented by different programmers, plenty of applications are able to develop by traditional software programmers on CPUs.

As shown in Fig. 3, when developing applications on FPGA-based heterogeneous systems with the OpenHEC framework, two steps of task can be omitted, which are FPGA system constructing and device drivers development. These two steps are all in system level, which need users to have more knowledge about computer architecture and operating systems. Furthermore, FPGA reconfiguration is a built-in function provided by OpenHEC, which is a nontrivial task if handled by application programmers.

## IV. EVALUATION

In order to show usability of OpenHEC for C/C++ application programmers, in this section, the implementing result of SURF feature detection based on OpenHEC is introduced.

A board for vision computing designed by our group is used to be the experimental platform. Of course, heterogeneous platforms employing SoCs can also be used to port the OpenHEC framework. As shown in Fig. 4, this board contains a Xilinx Spartan-6 XC6SLX150T FPGA and a 1GHz Samsung Cortex-A8 S5PV210 processor. The FPGA and the ARM processor are connected through 16-bit data bus and 16-bit address bus, which can be switched to 2 different address spaces by chip selection signals. 4 GPIOs ( General Purpose Input/Output) are used as control signals. 4-bit SPIs (Serial Peripheral Interface) and 7-bit GPIOs are used for the processor to configure bitstream for FPGAs. There are four 16-bit DDR3 connected with the FPGA to provide high bandwidth memory accessing for computer vision algorithms. CMOS and CameraLink interfaces are provided for video input.

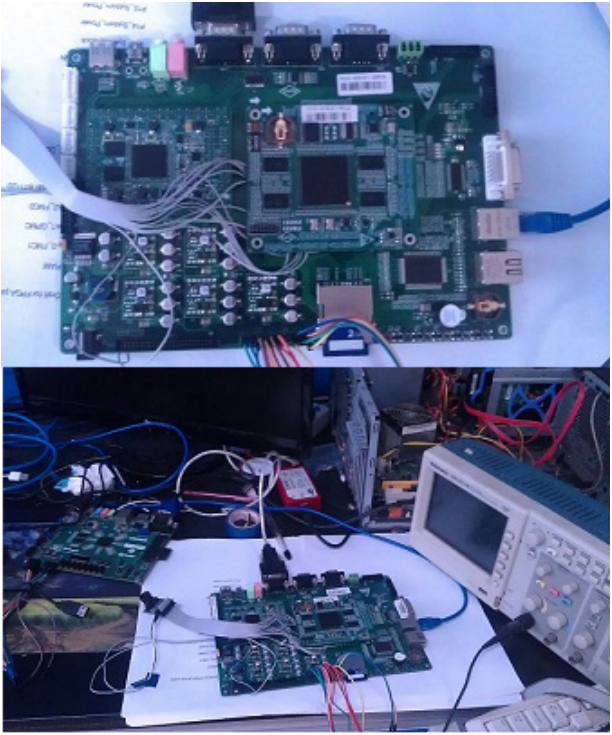

Fig. 4. ARM-FPGA Heterogenous Board for Vision Computing

### A. SURF Overview

SURF [16] is a widely-used algorithm for local feature detection and description. It is insensitive to different transformations, such as image scale, rotation and illumination. The SURF algorithm can be mainly divided into three stages: initialization, feature detection and feature description.

•**Initialization:** This stage does some initialization work such as loading the image, getting the intensity of each image pixel and calculating the intermediate representation known as integral image.

•**Feature detection:** This stage detects all the interest points in an image or a video frame. A Hessian based scale-space pyramid is first constructed and analyzed through scaling the original image several times. Therefore, the interest points can be selected from different scales of an image, which guarantees the algorithm is scale insensitive. After the scale space analysis, a $3\times3\times3$ non maximum suppression is applied to localize the most stable points with high contrast. That is, if the value of one point is larger than any other 26 points in its cube, it is extracted as an interest point.

•**Feature description:** In this stage, each interest point is assigned with a feature vector to describe the point. Firstly, a characteristic orientation is calculated to make the algorithm rotation invariant. Then, a descriptor window is constructed according to the orientation to calculate the feature vector. Finally, each interest point is represented by a 64-dimension vector in SURF, which is normalized to keep illumination invariant.

### B. Implementing SURF Detection on OpenHEC

Because the initialization and detection stages have clear parallelism [17], which are easier to be optimized by software programmers. Thus, in this paper, only initialization and detection stages are implemented on the FPGA. According to the design flow introduced in Sec. III, the design flow of implementing SURF based on OpenHEC is described as follows.

•**Design SURF detection code in C/C++.** Firstly, a HLS project is created in Xilinx Vivado. Then, to define the entry as:
void SURF_detect($volatile hls\_int32 * inout\_pix$,
$unsigned int byte\_rdoffset$,
$unsigned int byte\_wroffset$,
$int rows, int cols$).
The template of the entry is provided in OpenHEC for programmers convenience. Where $hls\_int32$ is the type for pixels that is 32-bit integer. $inout\_pix$ is the buffer for pixels input and output. $byte\_rdoffset$ is the offset to read pixels from and $byte\_offeset$ is the offset to write pixels to. $rows$ and $cols$ are size of the image. Finally, C/C++ code of SURF detection is implemented.

•**Make SURF detection code AXI4-compliant.** The AXI4-compliant interface for the SURF detection module can be defined by using utilities provided by the Vivado HLS as follows.

#pragma HLS interface ap_bus port = inout_pix depth = 1024

#pragma HLS resource core=AXI4M variable=inout_pix

#pragma HLS interface ap_ctrl_hs port = return register

#pragma HLS interface ap_none register port = byte_rdoffset

#pragma HLS interface ap_none register port = byte_wroffset

#pragma HLS interface ap_none register port = rows

#pragma HLS interface ap_none register port = cols

Below defines return register for AXI4-Lite.

#pragma HLS resource core=AXI4LiteS metadata="-bus_bundle AXI4_Lite_Slave" variable=return

#pragma HLS resource core=AXI4LiteS metadata="-bus_bundle AXI4_Lite_Slave" variable=byte_rdoffset

#pragma HLS resource core=AXI4LiteS metadata="-bus_bundle AXI4_Lite_Slave" variable=byte_wroffset

#pragma HLS resource core=AXI4LiteS metadata="-bus_bundle AXI4_Lite_Slave" variable=rows

#pragma HLS resource core=AXI4LiteS metadata="-bus_bundle AXI4_Lite_Slave" variable=cols

•**Optimize SURF detection code for Better Parallelism.** Software programmers can use optimizing utilities provided in Vivado HLS to do optimization for their C/C++ code. Further optimization can be carried out manually.

#pragma HLS DATAFLOW

#pragma HLS PIPELINE

●**Generate Pcore of SURF detection code.** To generate Pcores for EDK of SURF detection code in Vivado.

●**Mount Pcore to the FPGA Part of OpenHEC.** This step will connect the Pcore generated above with the FPGA part of OpenHEC. Due to the Spartan-6 FPGA is used in our development board, the Xilinx ISE will be used here. The OpenHEC framework on FPGAs is opened as a existing project. The SURF Pcore can be selected as an IP module, which can be connected easily with OpenHEC at a fixed port.

●**Generate Bitstream for Configuring the FPGA from the Processor and Design API for Upper Users on CPUs.** After fixing user's Pcore with the OpenHEC framework on FPGAs, the whole system can be synthesized to generate a bitstream file used by the processor to configure the FPGA. Then, an API is constructed according to introduction in Sec. III for upper software users on the CPU.

●**Design modules on CPUs based on the API above.** The last step will design modules executed on CPUs based on the API above to construct the whole application on the heterogeneous platform.

### C. Experimental Results

Firstly, execution time on the FPGA-based heterogeneous platform equipped with OpenHEC, the ARM-based embedded computer and the desktop computer respectively is displayed with comparison. Secondly, FPGA resources utilization of the SURF detection with and without OpenHEC is displayed. Thirdly, detection results for images are displayed to show functional correctness of this implementation and OpenHEC.

Besides the board introduced above, in this experiment, the working frequency of SURF detection module on FPGAs is 62.5MHz. The desktop CPU is 3.1GHz AMD Athlon(tm) II X4 645, with DDR3 memory working at 1.6GHz. The configuration of their parameters for the SURF detection is shown in Table I. Please note that the floating-point number is used in CPU based computers but fixed-point in FPGAs.

TABLE I.   PARAMETERS CONFIGURATION

| Parameter | Values |
|---|---|
| MinHessiam | 10 |
| nOctaves | 1 |
| Intervals | 2 |
| Pyramidal Level | 4 |

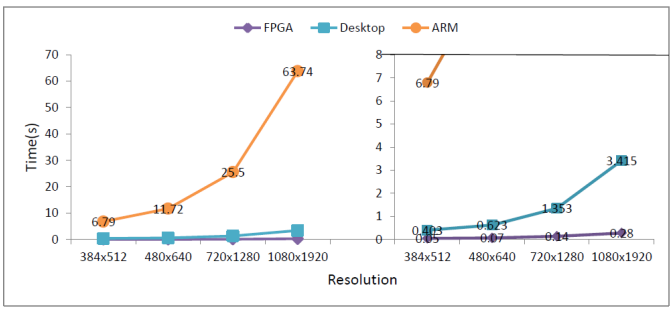

Fig. 5.   Execution Time of SURF Detection on Different Platforms

*1) Execution Time:* To make execution time of the SURF detection on different platforms comparable, the same group of images is used. Fig. 5 shows execution time of the SURF detection on different platforms with different image resolution. The right part of Fig. 5 shows the close-up for execution time on the FPGA and the desktop for clarity. As shown in Fig. 5, the SURF detection on the FPGA with OpenHEC is fastest and with the image increasing, the speed-up ratio is higher. It is because parallel computing makes time increased more linearly. It highlights significance to facilitate software programmers to use FPGAs.

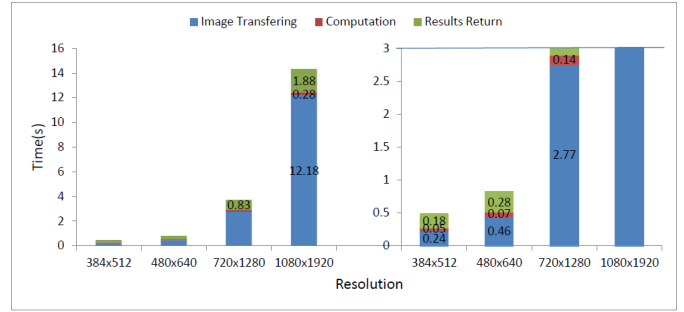

Fig. 6.   Execution Time of SURF Detection Including Data Transfering

Fig. 6 shows execution time of the SURF detection based on the FPGA-based heterogeneous platform that includes image transferring from ARM to FPGA, computing time and time to send results back to ARM. As shown in Fig. 6, to transfer image data from ARM to FPGA is most time-consuming. Although execution time including these three parts is still shorter than that of ARM, it should be avoided by capturing image data from FPGA camera interfaces directly. Furthermore, frequent reconfiguration should be avoided to save an around 2 s bitstream configuring time of the Spartan-6 XC6SLX150T.

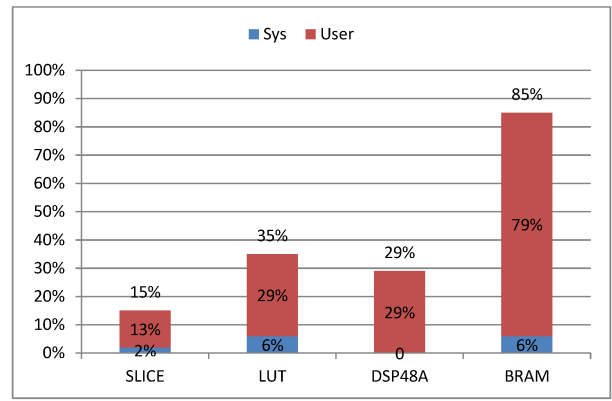

Fig. 7.   Resources Utilization of SURF Detection with and without OpenHEC

*2) Resources Utilization:* Fig. 7 shows resources utilization of SURF detection with and without the OpenHEC framework. As shown in this figure, a very small percentage of resources is consumed by OpenHEC. With integration degree of FPGAs increasing continuously, resources consumption by the system framework will be tolerated more willingly for ease of use.

*3) Detection Results:* Fig. 8 shows detection results of SURF detection on FPGA-accelerated heterogeneous platform

with OpenHEC. It shows functional correctness of this implementation and OpenHEC. Although fixed point employed in FPGA-accelerated version leads to accuracy differences to some extent, it can be improved through adjusting the threshold properly.

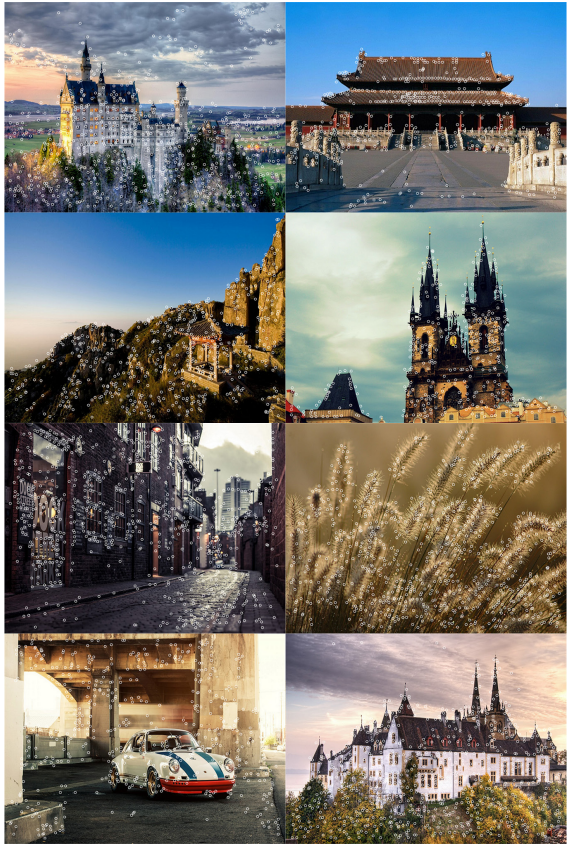

Fig. 8. Detection Results of SURF Detection on FPGA-accelerated Heterogeneous Platform

## V. CONCLUSION

Conventional development methods of FPGAs are far beyond the feild of software programmers, especially application programmers with little architectural knowledge. However, application programmers have best knowledge about requirements of markets and users. In the field of general processors, the multi-layer structure frees application programmers from low-level details, which improves productivity and decreases time to market. With integration degrees increasing continuously, FPGAs are also able to tolerate resources consumption coming from system virtualization. Furthermore, FPGAs also need virtualization technology to improve resources management and productivity. In this paper, a design framework for C/C++ application programmers to use FPGA-based heterogeneous platforms is proposed. This framework makes application users to focus more on implementing specific functions in C/C++ on FPGAs and CPUs respectively. The framework of OpenHEC was integrated into the commercial tool chain such as Xilinx Vivado to make it usable at once. Through implementing a popular feature detection system successfully on the FPGA-based heterogeneous platform based on this framework from a software perspective, it shows the OpenHEC framework is usable for application programmers. Next step, the debugging scheme suitable for software programmers is going to investigated and integrated into this framework. Furthermore, other virtualization techniques such as virtual memory, muti-FPGA partitioning etc been studied in RAMP, LEAP and CoRAM can also be integrated into this framework.

ACKNOWLEDGMENT

This work was supported by the Open Project Program of the State Key Laboratory of Mathematical Engineering and Advanced Computing.